\documentclass[twocolumn, secnumarabic, amssymb, noshowpacs,nobibnotes, aps, showpacs]{revtex4-1}
\usepackage{graphics}
\usepackage[dvips]{graphicx}
\usepackage{dcolumn}
\usepackage{amsmath}

\begin{document}

\title{Charge polarization effects on the optical response of blue-emitting superlattices}
\author{Pedro Pereyra$^a$, Fatna Assaoui$^b$}
\address{$^a$ Departamento de Ciencias B\'asicas, UAM-Azcapotzalco, M\'exico D.F., Mexico, C.P. 02200 \\
$^b$ Department of Physics, University Mohammed V, Av. Ibn Battouta, Rabat-Morocco, B. P. 1014}

\begin{abstract}

In the new approach to study the optical response of periodic structures, successfully applied to study the optical properties of blue-emitting InGaN/GaN superlattices, the spontaneous charge polarization was neglected. To search the effect of this quantum confined Stark phenomenon  we study the optical response, assuming parabolic band edge modulations in the conduction and valence bands. We discuss the consequences on the eigenfunction symmetries and the ensuing optical transition selection rules. Using the new approach in the WKB approximation of the finite periodic systems theory, we determine the energy eigenvalues, their corresponding eigenfunctions and the subband structures in the conduction and valence bands. We calculate the photoluminescence as a function of the charge localization strength, and compare with the experimental result. We show that for subbands close to the barrier edge the optical response and the surface states are sensitive to charge polarization strength.

\end{abstract}
\draft

\maketitle



\section{Introduction}
\begin{figure*}
\begin{center}
\includegraphics[width=340pt]{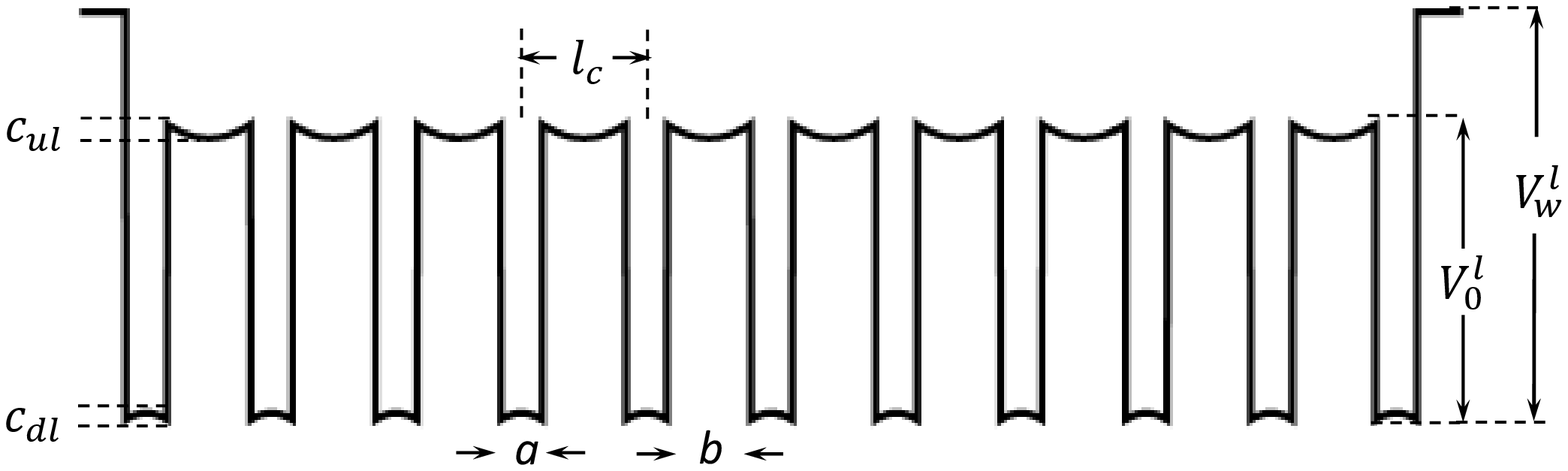}
\caption{Parabolic modulations of the potential profile and potential parameters at the conduction (l=c) and valence (l=v) bands of blue-emitting $InGaN$ superlattices, with spontaneous charge polarization at the valley-barrier interfaces.}
\label{figSL}
\end{center}
\end{figure*}
\label{sect:intro}  Recently, a new approach based on the theory of finite periodic systems, with explicit calculations of the emitter energy eigenvalues and eigenfunctions, was proposed to calculate the optical response of periodic structures\cite{PereyraAP2, SelecRules, SRLetter}. One of the examples, extensively discussed in these references, has been the high-resolution optical spectra of blue emitting InGaN superlatices (SLs), widely studied by Nakamura et al.\cite{NakamuraPaper} Excellent agreement with the experimental results was found assuming sectionally constant periodic potentials at the conduction and valence band edges. It is well known, however, that due to charge polarization at the superlattice layers' interfaces, the potential profiles, in the conduction and valence band edges, become parabolic\cite{Kosodoy, Nomura}, as shown in figure \ref{figSL}, where the index  $l=c,v$, stands for conduction and valence band.

In Ref. \citenum{Assaoui}, the effects of charge polarization on the transmission probabilities and the resonant band structure of open $In_xGa_(1-x)N/In_yGa_(1-y)N$ SLs, were studied and one of the results obtained there was that, for energies just above the barrier, the subbands become highly asymmetric. On the other hand, the successful theoretically calculations in Refs.~\citenum{SelecRules} and \citenum{SRLetter}, for the blue emitting samples, implied precisely transitions between subbands that are  close to the barrier edges. It is then worth asking whether the charge polarization effect has or not any consequence on the theoretical results. To this purpose, we consider here specifically the parabolic modulation of the conduction and valence  band-edges and calculate the  optical response for the same sample studied in Refs.~\citenum{SelecRules} and \citenum{SRLetter}.

In section \ref{sec: model}  we outline the model in the WKB approximation and discuss, briefly, the effects of charge polarization on the eigenfunction symmetries and on the selection rules. In section \ref{sec: optical responses} we calculate the energy eigenvalues structure for the blue emitting SLs and the corresponding optical response, and compare with the experimental and the theoretical calculation in the absence of charge polarization. We end up with some conclusions.

\section{The model}\label{sec: model}

For the calculation of the optical response we use the well known golden rule

\begin{eqnarray}\label{susceptPL}
\chi_{\small PL}\!\!=\!\! \sum_{\nu,\nu'\!,\mu,\mu'}\!f_{eh}\frac{\displaystyle \Bigl{|}\int dz
[\varphi^{v}_{\mu',\nu'}(z)]^*\frac{\partial}{\partial
z}\varphi^{c}_{\mu,\nu}(z)\Bigr{|}^{2}}{(\hbar \omega-E_{\mu,\nu}^{c}-E_g+E_{\mu',\nu'}^{v}+E_B)^{2}+\Gamma^{2}}.\nonumber \\
\end{eqnarray}
\begin{figure*}
\begin{center}
\includegraphics[width=380pt]{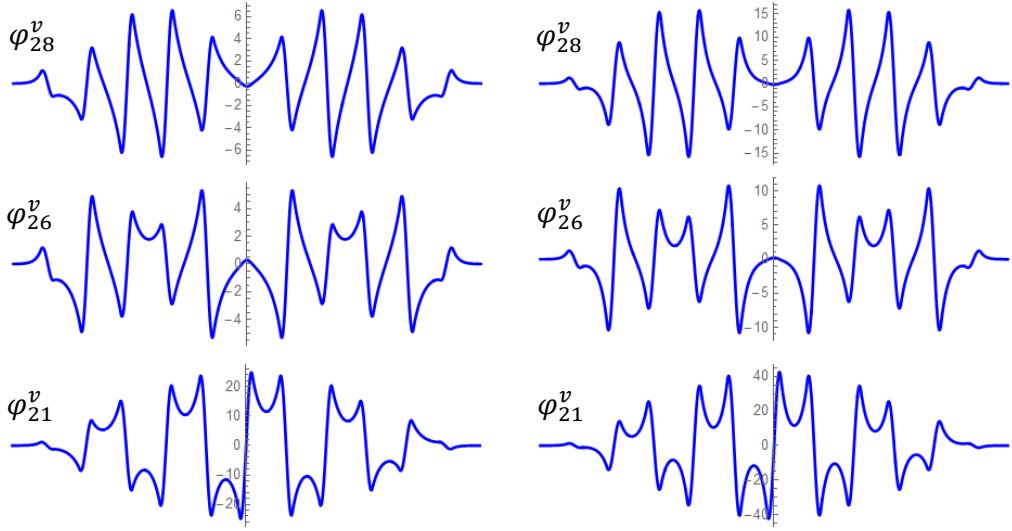}
\caption{Charge polarization effect on eigenfunctions. The eigenfunctions $\varphi_{2,1}^v$, $\varphi_{2,6}^v$, and $\varphi_{2,8}^v$, in the second subband of the valence band of the blue emitting $(In_xGa_{1-x}N/In_yGa_{1-y}N)^n In_xGa_{1-x}N$ superlattice bounded by $AlGaN$ clading layers, with (left) and without (right) charge polarization, $n=10$, $x=0.2$ and $y=0.05$.  }
\label{figEFs}
\end{center}
\end{figure*}
Here $\omega $ is the emitted photon frequency, $\Gamma$ the level broadening energy, $E_g$ the gap energy, $E_{\mu,\nu}^{c}$ and $E_{\mu',\nu'}^{v}$ are the energy eigenvalues in the conduction ($c$) and valence ($v$) bands, measured from the band-edges; $\mu$ and $\mu'$ denote the subband indices and $\nu$ and $\nu'$ the intra-subband energy levels. $\varphi^{c}_{\mu,\nu}(z)$ and $\varphi^{v}_{\mu,\nu}(z)$ the corresponding eigenfunctions and $f_{eh}$ the  occupation probability. $E_B$ is the exciton binding energy. All these quantities are explicitly and rigorously calculated within the effective mass approximation and the theory of finite periodic systems. Analytic and general formulas were derived and amply discussed in Ref. \citenum{PereyraAP1}. For the quasibound SLs considered here, with potential profiles as in figure \ref{figSL}, corresponding to potential functions $V_d^l(z)=-a_{dl}(z-z_0)^2+c_{dl}$ and $V_u^ l(z)=a_{ul}(z-z_0)^2+c_{ul}$,  in the valley and barrier of the conduction (l=c) and valence (l=v) bands, respectively, we use Eqs. (18) and (21-23) of Ref. \citenum{PereyraAP1}, written in the WKB approximation. This means that
quantities like  $k z$, $q z$, $ka$ and $q b$ must be replaced by (to simplify the notation we drop the band index $l$)
\begin{eqnarray}
K_z=\frac{1}{\hbar}\int_{z_r}^z dz \sqrt{2m^*\bigl(E+a_d(z-z_0)^2-c_d\bigr)},
\end{eqnarray}
\begin{eqnarray}
Q_z=\frac{1}{\hbar}\int_{z_r}^z dz \sqrt{2m^*\bigl(a_u(z-z_1)^2+c_u-E\bigr)},
\end{eqnarray}

\begin{eqnarray}
K_a&=&\frac{1}{\hbar}\int_0^a dz \sqrt{2m^*\bigl(E+a_d(z-z_0)^2-c_d)\bigr)}\cr &=&\frac{a}{\hbar}\sqrt{\frac{m^*}{2c_d}}\Bigl(\sqrt{c_d E}+(E-c_d)\tan^{-1}\sqrt{\frac{c_d}{E}}  \Bigr)
\end{eqnarray}
and
\begin{eqnarray}
Q_b&=&\frac{1}{\hbar}\int_{0}^b dz\sqrt{2m^*\bigl(a_u(z-z_1)^2+c_u-E)\bigr)} \cr &=&\frac{1}{\hbar}\sqrt{\frac{m^*} {a_u}}
\Bigl(b \sqrt{a_u (a_u b^2 + 4 c_u - 4 E)}\,\, + \cr &&
 4 (c_u - E) \tanh^{-1}\Bigl[\frac{a_u b}{\sqrt{
   a_u (a_u b^2 + 4 c_u - 4 E)}}\Bigr]\Bigr),\hspace{0.2in}
\end{eqnarray}
respectively, with $z_r$, $z_0$ and $z_1$ reference points, properly chosen.
\begin{figure*}
\begin{center}
\includegraphics[width=380pt]{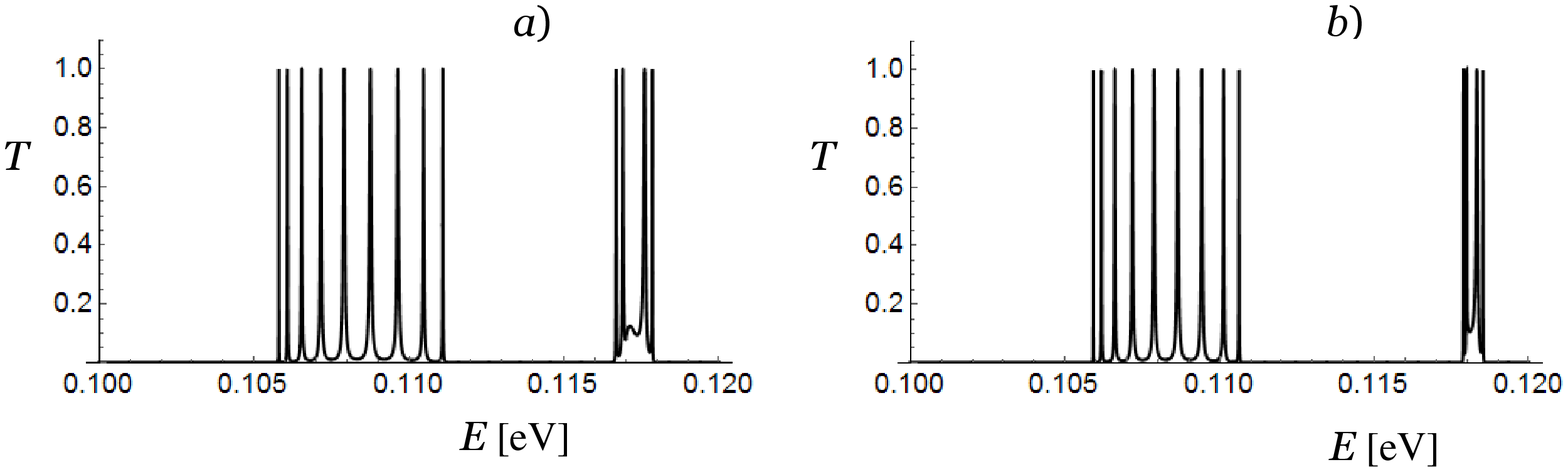}
\caption{Holes' transmission coefficients in open superlattices for two values of the parameter $c_{uv}$. In $a$) $c_{uv}=$3meV and $c_{uv}=$2meV in $b$).}
\label{figTCs}
\end{center}
\end{figure*}

\subsection{Charge polarization effects on symmetries and selection rules}

\begin{figure*}
\begin{center}
\includegraphics[width=400pt]{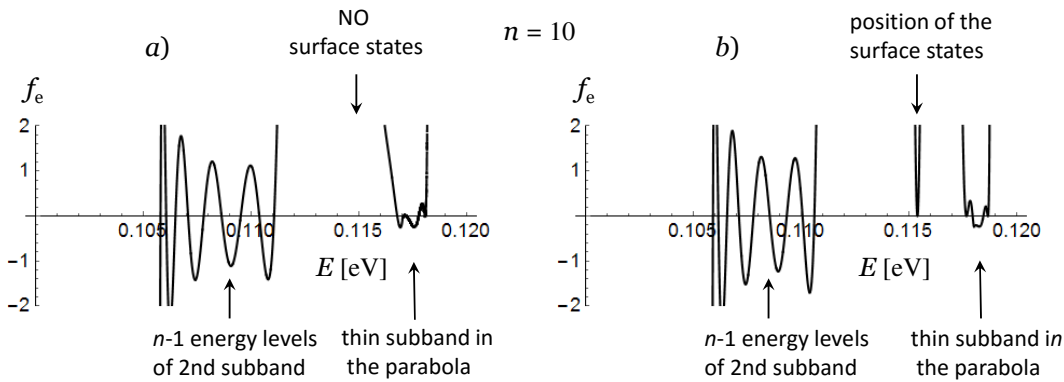}
\caption{The eigenvalues function $f_e(E)$  in the valence band for two values of $c_{uv}$. In $a$) $c_{uv}=$3meV and $c_{uv}=$2meV in $b$). In $b$), around 0.115eV, the function $f_e(E)$ crosses the energy axes, implying the existence of two surface energy levels. In this case $E_{2,10}$=0.1153773eV and $E_{2,11}$=0.1153777eV. Notice also the asymmetry in the subbands density of states. }
\label{figTrs}
\end{center}
\end{figure*}

Since the SL global spatial symmetry does not change because of the local parabolic modulation, it is clear that the eigenfunction parity symmetries, summarized in Ref. \citenum{PereyraAP2} as
\begin{eqnarray}\label{QBWFSym}
\Psi_{\mu,\nu}(z)\!=\!\Biggl\{\begin{array}{cc} (-1)^{\nu+1}\Psi_{\mu,\nu}(-z) & {\rm for}\hspace{0.1in} n \hspace{0.1in}{\rm odd}\cr  (-1)^{\nu+\mu}\Psi_{\mu,\nu}(-z) & \hspace{0.05in}{\rm for}\hspace{0.1in} n \hspace{0.1in}{\rm even},  \end{array}\Biggr.
\end{eqnarray}
remain unchanged. This is apparent in figure \ref{figEFs}, where some eigenfunctions for the superlattice with parabolic modulation (left) are plotted together with the eigenfunctions for the superlattice with sectionally constant (right) potential profile. Consequently, the symmetry selection rules, that rely on the eigenfunction symmetries, remain the same as proposed in Ref. \citenum{SelecRules}.

At variance with the  eigenfunctions behavior, the energy eigenvalues and the surface states are sensitive to the strength of the charge polarization, represented in our model  by the heights ($c_{dc}$ and $c_{dv}$ ) and depths ($c_{uc}$ and $c_{uv}$) of the parabolic modulations. In general, the conduction band edge modulation pushes up the energy eigenvalues. In some cases this effect may have no other consequence than a shift of the optical spectra, but in others we can have important changes in the energy-level structure. To gain an insight into this effect, we plot in figure \ref{figTCs} the resonant {\it hole's} transmission coefficients, for two values of the parameter $c_{uv}$. In $a$), we considered $c_{uv}=$3meV while in $b$) $c_{uv}=$2meV. The other parameters were exactly the same for $a$) and $b$).  It is well-known that the resonant bands, of open SLs, provide good information on the position of the eigenvalues bands of bounded SLs. As mentioned before the optical transitions that account for the observed spectra  correspond to those from the first subband in the conduction band, with energies of the order of 0.13eV (for a barrier height $V_{0}^c\sim$ 0.24eV), to the second subband of the valence band (VB), with barrier height  $V_{0}^v\sim$ 0.12eV, and energy eigenvalues spread out around 0.11eV. Notice that this subband (the second of the VB) appears in figure \ref{figTCs} just below the barrier edge, while four resonances appear for energies within the parabolic confining potential. Notice also that the subbands in $a$) are slightly wider than in $b$).

\begin{figure*}
\begin{center}
\includegraphics[width=380pt]{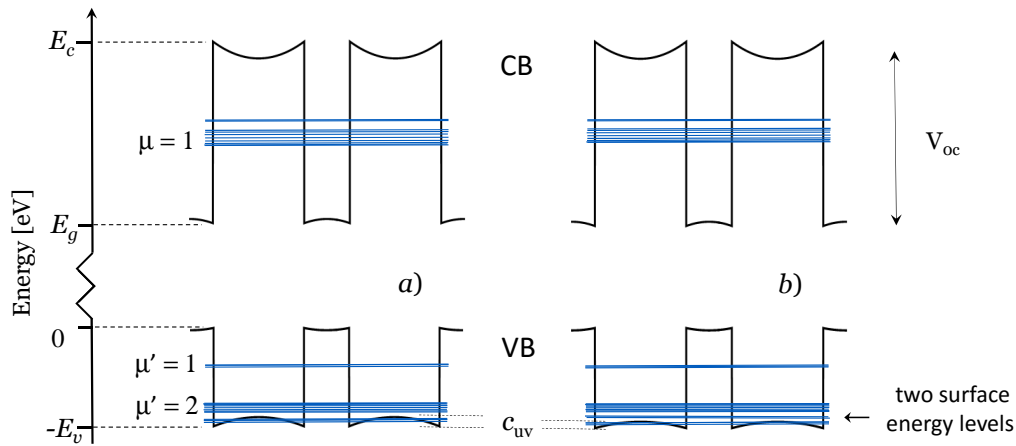}
\caption{The subbands in the conduction and  the valence bands for two values of $c_{uv}$. In $a$) $c_{uv}=$3meV and in $b$) $c_{uv}=$2meV. This small change has a large effect on the energy levels structure close to the valence band barrier edge. The most important is the presence of the surface energy levels in b), that are absent in a).}
\label{figCvcs}
\end{center}
\end{figure*}
Since the change in the parameter $c_u$ is small, it is convenient to see the effect on the band structure through the eigenvalues equation
\begin{eqnarray}
f_e(E)&=&\Re{\rm e} [\alpha_{nl}e^{ik_la}]-\frac{k_l^2-q_{wl}^2}{2k_lq_{wl}}\Im{\rm m} [\alpha_{nl}e^{ik_la}]\,\,\cr && - \frac{k_l^2+q_{wl}^2}{2k_lq_{wl}}\Im{\rm m}[\beta_{nl}]=0.
\end{eqnarray}
Where $\alpha_{nl}$ and $\beta_{nl}$ are the elements (1,1) and (1,2) of the SL transfer matrix $M_{SL}^l$, and $k_l$ and $q_{wl}$ the wave numbers
\begin{eqnarray}
k_l=\sqrt{\frac{2m^*_l E}{\hbar^2}}\hspace{0.2in} {\rm and}\hspace{0.2in}q_{wl}=\sqrt{\frac{2m^*_l (V_{w}^l-E)}{\hbar^2}},
\end{eqnarray}

at the wells and cladding layers of the conduction (l=c) and valence (l=v) b positions ands. It is clear that plotting $f_e(E)$ we can easily visualize the energy eigenvalues distribution. It is well known that for bounded superlattice with $n$ unit cells, each subband contains $n+1$ energy levels.\cite{PereyraCastillo} Two of them correspond to surface states and detach from the remaining $n$-1. When the cladding-layer barriers are both of the same height the surface states are practically degenerate. In figure \ref{figTrs} we plot the eigenvalues function in the VB for $c_{uv}=$3meV, in $a$), and for $c_{uv}=$2meV, in $b$). Notice that in $b$) the  function $f_e(E)$, around 0.115eV, approaches and touches the  energy axes. This behavior, implies the existence of additional energy levels. In this case the surface states, that are absent in $a$).
\begin{figure*}
\begin{center}
\includegraphics[width=380pt]{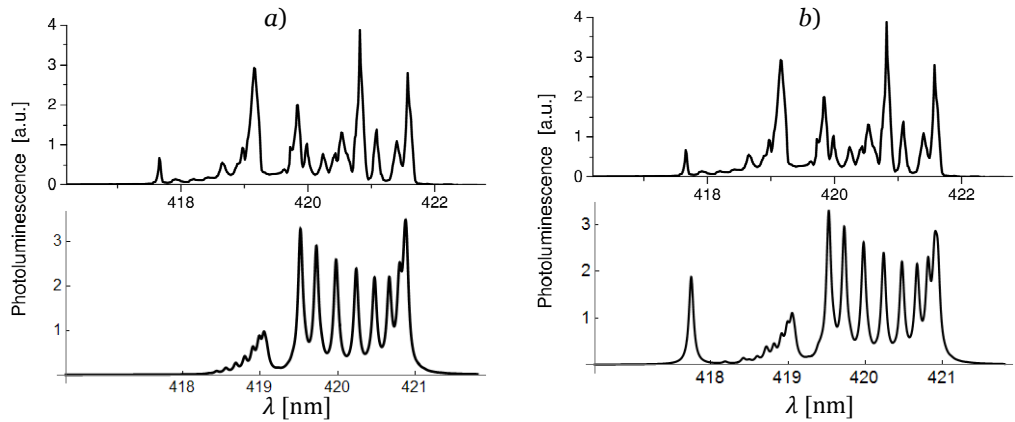}
\caption{The photoluminescence of the blue emitting $(In_xGa_{1-x}N/In_yGa_{1-y}N)^n In_xGa_{1-x}N$ superlattice. In the upper panel the experimental results. In the lower panels the theoretical calculations  for two values of $c_{uv}$. In $a$) $c_{uv}=$3meV and in $b$) $c_{uv}=$2meV. We have a better agreement in $b$) than in $a$). The experimental spectrum is reproduced with permission
from [{\it Appl. Phys. Lett.} {\bf 68}, 3269 (1996)]. Copyright [1996], AIP Publishing LLC.}
\label{figPLs}
\end{center}
\end{figure*}

In figure \ref{figCvcs}, we plot the subbands in the conduction and the valence bands. The subband for $\mu'$=2, in the VB, is slightly amplified to visualize the effect. Besides the subbands with and without surface-energy levels, we see also the new energy levels in the confining and shallow parabolas at the barrier edge, both in \ref{figCvcs}$a$) and \ref{figCvcs}$b$).

In the next section we will see the consequences of the presence or absence of the surface states on the optical spectra.

\section{The optical responses}\label{sec: optical responses}

We shall now present the charge polarization effect on the optical spectra of the blue emitting $(In_xGa_{1-x}N/In_yGa_{1-y}N)^n In_xGa_{1-x}N$ superlattice bounded by $AlGaN$ layers, for $n=10$, $x=0.2$ and $y=0.05$. The photoluminescence spectra shown in the lower panels of figures \ref{figPLs}$a$) and \ref{figPLs}$b$) were calculated using the golden rule in Eq. (\ref{susceptPL}). The energy eigenvalues and eigenfunctions were obtained from the theory of finite periodic systems in Ref. \citenum{PereyraAP1} and the symmetries and ensuing selection rules from Refs. \citenum{PereyraAP1} and \citenum{SelecRules}. The optical responses were calculated for the two values of $c_u$ discussed in section \ref{sec: model}. The spectrum in \ref{figPLs}$a$) is for $c_{uv}=$3meV and the spectrum in \ref{figPLs}$b$) for $c_{uv}=$2meV. In the upper part of these figures we show also the experimental result in Refs. \citenum{NakamuraPaper} and \citenum{NakamuraBook}. It is clear from this figures that a better agreement is found when $c_{uv}=$2meV. For slightly larger value of the parameter $c_{uv}$, that corresponds to a stronger charge polarization strength, we miss the surface energy levels responsible for the optical transitions at $\lambda \sim$417.7nm. As explained amply in Refs. \citenum{SelecRules} and \citenum{SRLetter}, the resonances in the lower panel of \ref{figPLs}$a$) appear in two groups because of the detachment of the surface states, as indicated with arrows in \ref{figCvcs}$a$). The transition from the surface state in the CB to the surface state in the VB appears in \ref{figPLs}$b$) as an isolated peak.

To plot the PL spectra in figure \ref{figPLs} we considered fixed values for $E_g$ and $E_B$, such that $E_g-E_B$= 2.716eV; the unit-cell length $l_c$= 7.3nm and $\Gamma$=0.00025eV. The predicted peak separations are of the order of 0.15meV equivalent to $\sim$0.2nm in full agreement with the experimental results. This separation corresponds with the intrasubband energy eigenvalues separation, as was glimpsed in Ref. \citenum{NakamuraBook}, after showing that the observed peak separations can `` NOT (be attributed) to simple Fabry-Perot modes", see page 268 of Ref. \citenum{NakamuraBook}. The experimental measurements were obtained with a resolution of 0.016nm.

To conclude this letter, it is worth stressing that the charge polarization corrections discussed here may, in some cases, be necessary to perform in order to account for particular optical spectra features. Our calculations  confirm also the relevance of the new theoretical approach to study the optical response of periodic structures. The high accuracy of the theoretical model makes it a powerful and simple approach to design laser devices for different purposes, including laser devices of interest in health applications.

\section{Conclusions}\label{sec: conclusions}

We have shown here that the parabolic modulation of the valley and barrier band edges, in the conduction and valence bands, have no effect on the eigenfunction symmetries and the selection rules. We have shown with explicit calculations for the blue-emitting $(In_xGa_{1-x}N/In_yGa_{1-y}N)^n In_xGa_{1-x}N$ superlattice bounded by $AlGaN$ layers,  that the charge polarization strength may have effects on the energy-eigenvalues structure close to the barrier edge. A theoretical calculation, as the one presented here, can not only account for the observed high resolution spectra, it can also provide an insight into the spontaneous charge polarization strength.


\begin{thebibliography}{99}

\bibitem{PereyraAP2}Pereyra P., ``Theory of finite periodic systems: The eigenfunctions symmetries," {\it Ann. Phys.} {\bf 378},264 (2017).
\bibitem{SelecRules}Pereyra P., ``Improved optical transitions theory for superlattices and periodic systems; new selection rules", arxiv condmatt 1607.02686.
\bibitem{SRLetter}Pereyra P., ``New approach to study light-emission of periodic structures. Unveiling novel surface-states effects", arXiv condmatt 1612.09002.
\bibitem{NakamuraPaper} Nakamura S., Senoh M., Nagahama S., Iwasa N., Yamada Ta., Matsushita T., ``Characteristics of $InGaN$ multi-quantum-well-structure laser diodes", {\it App. Phys. Lett.} {\bf 38}, 3269-3271 (1996).
\bibitem{Kosodoy}Kozodoy P., Hansen M., Denbaars S. P., and Mishra U. K.,``Enhanced Mg doping efficiency in $Al_{0.2}Ga_{0.8}N/GaN$ superlattices" Appl. Phys.
Lett. 74, 3681 (1999).
\bibitem{Nomura} Nomura M.,Arakawa Y., Shimura T. and Kuroda K., ``Optical Control of Transmittance by Photo-Induced Absorption Effect in InGaN/GaN Structures",Jp. J. of Appl. Phys. {\bf 44}, 7238–7243, (2005).
\bibitem{Assaoui} Assaoui F. and Pereyra P., ``Charge polarization effects and hole spectra characteristics in $Al_xGa_{1-x}N\backslash GaN$ superlattices", {\it J. Appl. Phys.} {\bf 91}, 5163-5169 (2002).
\bibitem{PereyraAP1}Pereyra P., ``Eigenvalues, eigenfunctions, and surface states in finite periodic systems," {\it Ann. Phys.} {\bf 320}, 1-20 (2005).
\bibitem{NakamuraBook} Nakamura S., Pearton S. and Fasol G., {\it The Blue Laser Diode. The complete history} (Springer-Verlag, Berlin Heidelberg 1997 ). See pages 247, 268.
\bibitem{PereyraCastillo}Pereyra P. and Castillo E., ``Theory of finite periodic systems: General expressions and various simple and illustrative examples", {\it Phys. Rev. B} {\bf 65}, 205120-26 (2002).


\end{thebibliography}
\end{document}